\documentclass[aps,prb,reprint,superscriptaddress]{revtex4-1}

\usepackage[pdftex]{graphicx}
\usepackage{color}
\usepackage{enumitem}

\begin{document}


\title{Extended skyrmion lattice scattering and long-time memory in the chiral magnet Fe$_{1-x}$Co$_x$Si}



\author{L.J. Bannenberg}
\affiliation{Faculty of Applied Sciences, Delft University of Technology, Mekelweg 15, 2629 JB Delft, The Netherlands}
\author{K. Kakurai}
\affiliation{Neutron Science and Technology Center, CROSS Tokai, Ibaraki 319-1106, Japan}
\affiliation{RIKEN Center for Emergent Matter Science (CEMS), Wako 351-0198, Japan}
\author{F.Qian}
\affiliation{Faculty of Applied Sciences, Delft University of Technology, Mekelweg 15, 2629 JB Delft, The Netherlands}
\author{E. Leli\`{e}vre-Berna}
\affiliation{Institut Laue-Langevin, 71 Avenue des Martyrs, CS 20156 Grenoble, France}
\author{C.D. Dewhurst}
\affiliation{Institut Laue-Langevin, 71 Avenue des Martyrs, CS 20156 Grenoble, France}
\author{Y. Onose}
\affiliation{Department of Basic Science, University of Tokyo, Tokyo, 153-8902, Japan}
\author{Y. Endoh}
\affiliation{RIKEN Center for Emergent Matter Science (CEMS), Wako 351-0198, Japan}
\author{Y. Tokura}
\affiliation{RIKEN Center for Emergent Matter Science (CEMS), Wako 351-0198, Japan}
\affiliation{Department of Applied Physics, University of Tokyo, Tokyo 113-8656, Japan}
\author{C. Pappas}
\affiliation{Faculty of Applied Sciences, Delft University of Technology, Mekelweg 15, 2629 JB Delft, The Netherlands}

\date{\today}

\begin{abstract}
Small angle neutron scattering measurements on a bulk single crystal of the doped chiral magnet Fe$_{1-x}$Co$_x$Si with $x$=0.3 reveal a pronounced effect of the magnetic history and cooling rates on the magnetic phase diagram. The extracted phase diagrams are qualitatively different for zero and field cooling and reveal a metastable skyrmion lattice phase outside the $A$-phase for the latter case. These thermodynamically metastable skyrmion lattice correlations coexist with the conical phase and can be enhanced by increasing the cooling rate. They appear in a wide region of the phase diagram at temperatures below the $A$-phase but also at fields considerably smaller or higher than the fields required to stabilize the $A$-phase. 
\end{abstract}
\pacs{75.10.-b
75.30.Kz
75.25.-j	
}

\maketitle
\section{\label{sec:level1} Introduction}
Spin chirality generated by Dzyaloshinsky-Moriya (DM) interactions \cite{D, M}  is  the focus of interest due to the emergence of  chiral skyrmions \cite{bogdanov1989,bogdanov1994, Rossler:2006cq, muhlbauer2009, yu2010, munzer2010, nagaosa2013}, which are non-coplanar and topologically stable spin textures. These can form a unique type of long-range magnetic order, a skyrmion lattice (SkL),  as observed in  the isostructural B20 transition-metal silicides, TMSi (TM=Mn, Fe, Co), and germanides like FeGe by neutron scattering \cite{muhlbauer2009, munzer2010, Moskvin:2013kf}  and in real space by Lorentz transmission microscopy \cite{yu2010, Yu2011FeGe}. In these bulk cubic helimagnets, the SkL correlations appear spontaneously in the so called A-Phase, a small pocket in the magnetic field ($B$), temperature ($T$) phase diagram slightly below the transition temperature $T_C$ \cite{muhlbauer2009,yu2010}. In confined geometries such as thin films \cite{Yu2011FeGe} or nanowires \cite{Du:2014ek}, this narrow pocket expands and tends to cover a substantial part of the phase diagram below $T_C$ up to the lowest temperature.

Recent findings show that it is also possible in bulk MnSi to quench the thermodynamically stable SkL into a metastable state by  rapid cooling down to low temperatures \cite{oike2016}. Additionally, short range order that may be associated with isolated skyrmions has been found outside the A-Phase in MnSi \cite{Grigoriev:2014fa}. Thus the experimentally observed stability limits of chiral skyrmions and SkL in the reference cubic helimagnets seem to be less well defined and established than assumed so far. On the other hand, it is theoretically  established that metastable SkL and single skyrmions should exist over a broad range of the phase diagram \cite{bogdanov1994, Wilson:2014gd}, as supported by recent findings on thin films \cite{Romming:2015il} or on the bulk polar magnetic semiconductor $\text{GaV}_4\text{S}_8$ \cite{Kezsmarki:2015bw}. 

The  small angle neutron scattering (SANS) results presented below go further in this direction and show strong memory effects and very weak patterns with the characteristic SkL sixfold symmetry that coexist with the conical phase, indicating the stabilization of SkL outside the usual thermodynamic equilibrium limits in the bulk cubic helimagnet Fe$_{1-x}$Co$_x$Si, $x$=0.3. The sample belongs to the semi-conducting system Fe$_{1-x}$Co$_x$Si, which  is characterized by very long helix periods from $\ell \sim25$~nm to $300$~nm \cite{Beille:1981dx, 1985JPSJ...54.2975I} and by a change of magnetic chirality from left to right handed as the amount of Co doping increases, triggered by the change of the chemical lattice chirality at $x=0.2$\cite{2009PhRvL.102c7204G}. Fe$_{0.7}$Co$_{0.3}$Si has a right handed, or clockwise, chirality and the helices propagate along the [100] crystallographic directions. 
\begin{figure*}[tb]
\includegraphics[width=1\linewidth]{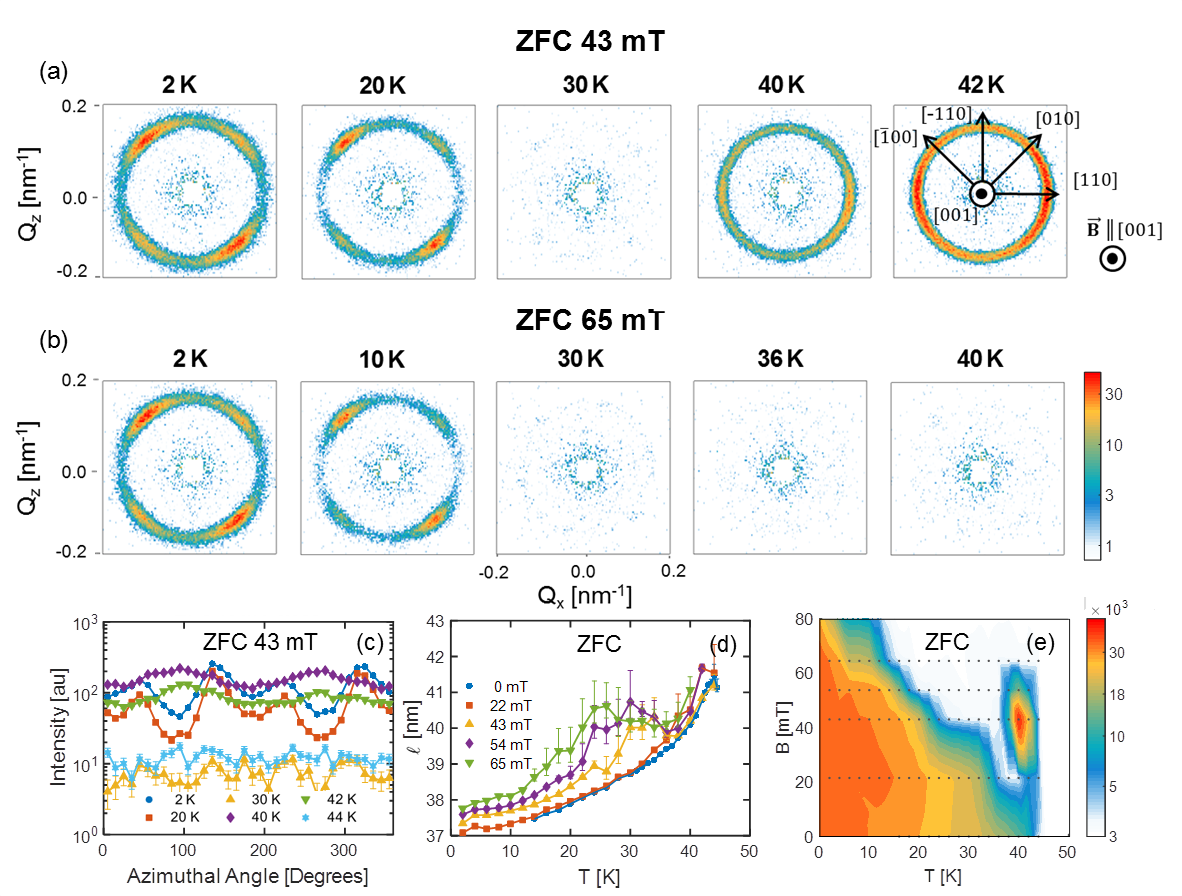}
\caption{\label{Fig:ZFC} SANS results obtained in ZFC configuration. Characteristic patterns at 43~mT and 65~mT are shown in panels (a) and (b).  The azimuthal average of the SANS intensity at 43~mT is given in (c) for selected temperatures. The temperature dependence of the  helical modulation period $\ell$ is given in (d) for the magnetic fields indicated. The deduced magnetic field and temperature dependence of the  total scattered intensity is given as a contour plot (e) and the dashed lines mark the magnetic fields corresponding to the patterns of (a) and (b).} 
\end{figure*}

Similar to chiral magnets of the same family, the ground state results from the competition between three terms in the Hamiltonian: a strong ferromagnetic exchange, a weaker DM  interaction and a weakest anisotropy\cite{bak1980}.  Below the transition temperature $T_C$ a helical order sets in with  $\ell$  proportional to the ratio of ferromagnetic exchange to DM interactions and with the helices fixed to the chemical lattice by anisotropy. This hierarchy is also found in the $B-T$ phase diagram, where a weak critical field $B_{C1}$ is enough to overcome the anisotropy, unpin the helices from the chemical lattice and orient them along its direction leading to the conical phase. A  higher magnetic field $B_{C2}$ is subsequently required to overcome the DM interactions and ferromagnetically align the magnetic moments inducing the spin polarized phase.  The $A$-phase occurs in a narrow region below $T_C$ and for intermediate magnetic fields between $B_{C1}$  and  $B_{C2}$\cite{beille1983,ishimoto1995,grigoriev2007,takeda2009,munzer2010}. 

A phase diagram depending on the magnetic history has already been found in Fe$_{1-x}$Co$_{x}$Si by neutron scattering \cite{munzer2010} and specific heat or magnetic susceptibility measurements\cite{Bauer:2016vq}. We chose to systematically investigate this effect and cooled the sample through $T_C$  following three specific protocols:  Zero Field Cooling (ZFC) and slow or fast Field Cooling (FC). Our results show a pronounced history effects and the existence of SkL correlations over a very extended region of the phase diagram when applying field cooling. These SkL correlation can be enhanced by increasing the cooling rate and do not only exist at temperatures below the $A$-phase.
\begin{figure*}[tb]
\includegraphics[width=1\linewidth]{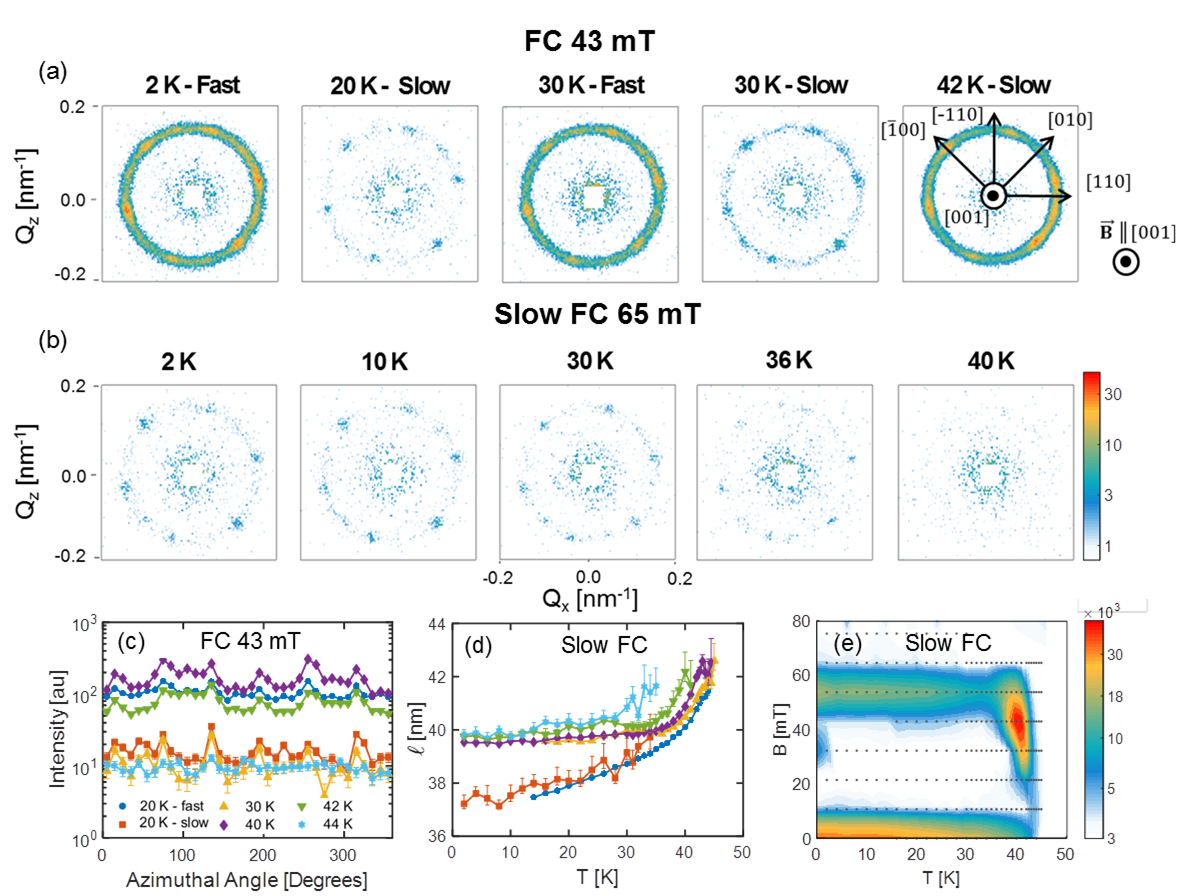}
\caption{\label{Fig:FC} SANS results obtained in FC configuration. Characteristic patterns at 43~mT and 65~mT are shown in panels (a) and (b). As mentioned in the text we differentiate between fast and slow field cooling. The azimuthal average of the SANS intensity at 43~mT is given in (c) for selected temperatures. The temperature dependence of the helical modulation period $\ell$ is given in (d) for slow FC and the magnetic fields indicated. The deduced slow FC magnetic field and temperature dependence of the  total scattered intensity is given as a contour plot (e) and the dashed lines mark the magnetic fields corresponding to the patterns of (a) and (b).}
\end{figure*}

\section{\label{sec:exp} Experimental Details}
The experiments were performed on the Small Angle Neutron Scattering (SANS) instrument D33 of the ILL using a monochromatic neutron beam with a wavelength  $\lambda$=0.6 nm and $\Delta\lambda / \lambda=10\%$ and on the same Fe$_{0.7}$Co$_{0.3}$Si single crystal ($\sim0.1\text{~cm}^3$) used in a previous investigation\cite{takeda2009}. The sample was oriented with the [$\bar{1}$10] axis vertical and the [001] axis parallel to $\vec{k}_i$,  the incoming neutron beam wavevector.  The data were normalized to standard monitor counts and a measurement at $T$=60 K was used for the background correction. The magnetic field $\vec{B}$ was applied parallel to $\vec{k}_i$ a  configuration, where only helical modulations that propagate perpendicularly to $\vec{B}$ may fulfill the Bragg condition and give rise to scattering. The results are thus complementary to the previous investigation\cite{takeda2009}, where the magnetic field was applied \textit{perpendicular} to the neutron beam and in the SANS detector plane. The magnetic field was applied following three specific protocols:  
\begin{itemize}
\item ZFC temperature scans: the sample was cooled down to 2~K under zero magnetic field, then a magnetic field was applied and the patterns were recorded by increasing the temperature in steps of 2~K every 6~min.
\item FC temperature scans: the magnetic field was applied at 60~K and the measurements were performed by decreasing the temperature in steps of 0.5~K every 10~min between 45~K and 42~K, in steps of 1~K every 5~min between 41~K and 30~K and between 30~K and 2~K in steps of 2~K every 5~min. 
\item Fast FC temperature scans: the magnetic fields of 43~mT or 54~mT were applied at 60~K and the sample was immediately brought to 30~K at a cooling rate of $\sim$ 3~K/min. The measurements were subsequently performed by decreasing the temperature to 2~K in steps of 2~K every 5~min. 
\end{itemize}

\section{\label{sec:results} Experimental Results}


\begin{figure*}[tb]
\includegraphics[width=1\linewidth]{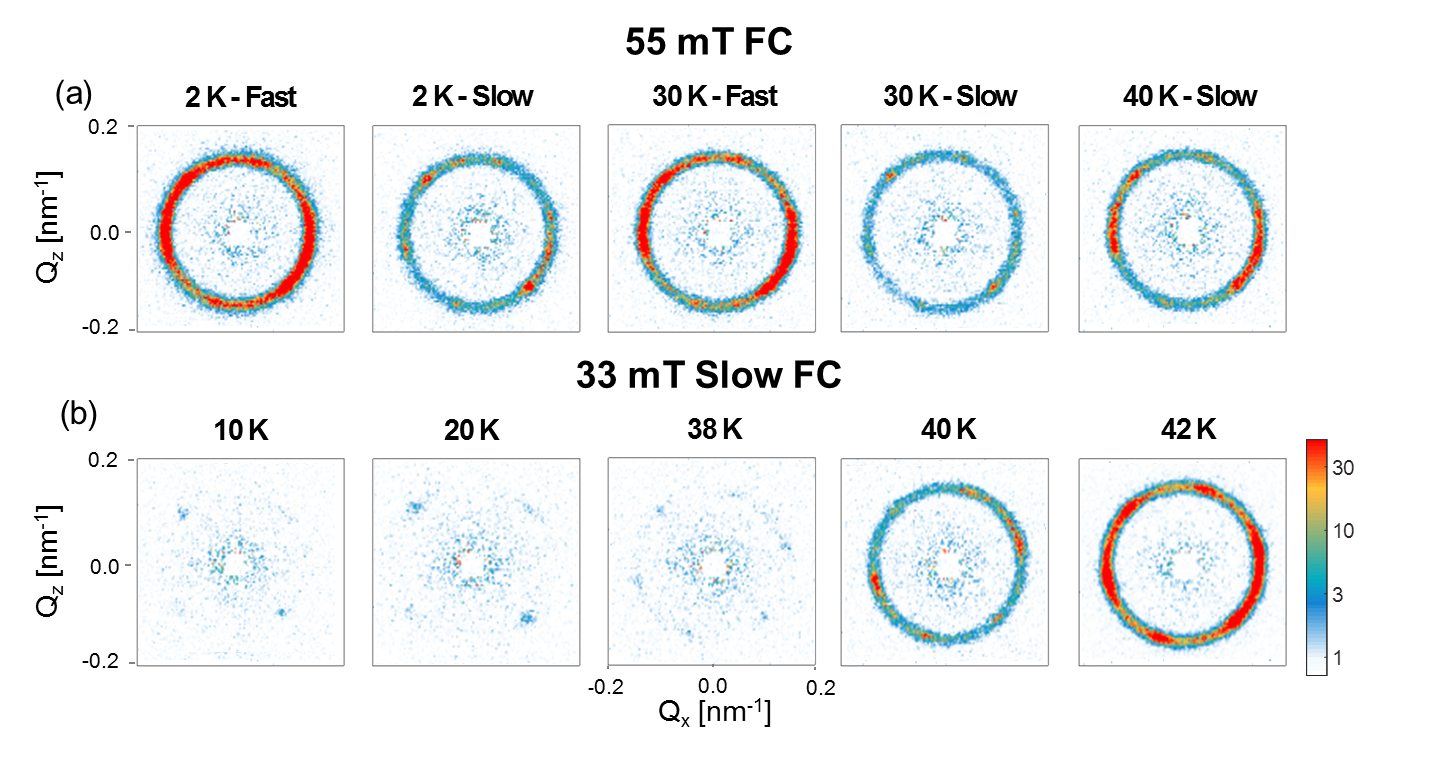}
\caption{Characteristic SANS patterns at (a) 33~mT and (b) 55~mT recorded for fast and slow field cooling.}
\label{Fig:Additional_Spectra} 
\end{figure*}


First, we present the ZFC results that are summarized in Fig. \ref{Fig:ZFC}, which shows typical SANS patterns for (a) $B=~$43~mT and (b) 65~mT, azimuthally averaged intensity at $B=~$43~mT (c), the temperature and magnetic field dependence of $\ell$ (d) and finally a $B-T$ contour plot of the integrated SANS intensity illustrating the occurrence of the different phases (e). The patterns at low temperatures displayed in panels (a) and (b) show four broad peaks that are the fingerprint of the helical order along the $\langle100\rangle$  crystallographic directions. By increasing the temperature the intensity almost vanishes at 30~K. This is the signature of the conical phase, where helices are aligned by the magnetic field and they do not fulfill the Bragg condition in the configuration of this experiment ($\vec{B} \|  \vec{k}_i$) and thus do not scatter neutrons. Additional information on this phase can be found in the previous work\cite{takeda2009}, where a complementary experimental set-up with $\vec{B}\perp \vec{k}_i$ was used. 

By further increasing the temperature the scattered intensity increases for $B$=43~mT and scattering patterns reappear for $T$ $\geq$ 38~K as shown in Fig. \ref{Fig:ZFC}(a) for $T$=40~K and 42~K. However, the observed pattern is not the six-fold SkL symmetry of MnSi \cite{muhlbauer2009} but a ring as illustrated by the azimuthally averaged intensities shown in panel (c). This behavior is similar to Fe$_{0.8}$Co$_{0.2}$Si where such a ring-like pattern was found and was attributed to the combination of disorder, arising from the solution of Fe and Co in the chemical lattice, and magneto-crystalline anisotropy\cite{munzer2010}. The azimuthal intensity plots also show that the four helical peaks visible at low temperatures do not have exactly the same intensity, reflecting a slight misalignment of the sample as pointed out in the previous investigation\cite{takeda2009}. 

Fig. \ref{Fig:ZFC}(d) depicts the temperature dependence period of the helical modulations $\ell$ that has been derived from the  momentum transfer $Q$ where the scattered intensity is maximum: $\ell=2\pi/Q$. As the temperature increases from 2 to 40~K, $\ell$ increases substantially by about 14 \%, which suggests a weakening of the DM interaction with respect to the ferromagnetic exchange. In addition for $B > 22$~mT, a non-monotonic temperature dependence is found with $\ell$ going through a minimum at the A-Phase and then through a maximum at a lower temperature which depends on the magnetic field.

FC results are summarized in Fig. \ref{Fig:FC}, which is complementary to Fig. \ref{Fig:ZFC}: the patterns in panels (a) and (b) are given for the same magnetic fields as for ZFC and reveal substantial  differences.  In contrast to ZFC, the helical phase is confined to magnetic fields below 10~mT. In the $A$-phase at 43~mT a ring-shaped scattering is found for 40~K and 42~K, shown in Fig. \ref{Fig:FC}(a), but in contrast to ZFC the six-fold symmetry characteristic of the SkL phase is visible in the FC patterns as well as in the corresponding azimuthal plots shown in panel (c). This six-fold symmetry pattern remains, although weak, clearly visible, even when the temperature is further decreased below 38~K, which suggests the coexistence of a weak SkL with the conical phase. SkL correlations therefore seem to freeze by cooling the sample in a magnetic field, and this effect depends on the cooling rate through the $A$-phase as shown by the different SANS patterns for slow and fast FC at 20~K and 30~K in Fig. \ref{Fig:FC}(a).  This freezing of the correlations is also seen in the evolution of $\ell$ in Fig. \ref{Fig:FC}(d), which for $B$ \textgreater 22~mT locks-in to the value at $\sim$38~K while cooling down, which is in sharp contrast with the ZFC behavior shown in Fig. \ref{Fig:ZFC}(d). 

For the slightly higher magnetic field of 55~mT the patterns of Fig. \ref{Fig:Additional_Spectra} show that the $A$-phase clearly extends down to 2~K. In this case, fast FC leads to higher scattered intensities than slow FC, although with the same symmetry and overall shape of the scattering patterns. The $B-T$ SANS intensity map in Fig. \ref{Fig:FC}(e) illustrates the boundaries of the $A$-phase, which extend to the lowest temperatures between 45 and 65~mT. However, outside this $A$-phase region, weak six-fold symmetry patterns appear with an intensity about 100 times weaker than in the $A$-phase, similarly to what is shown in Fig. \ref{Fig:FC} (a) and (b). In addition, at intermediate magnetic fields (10 \textless $B$ \textless 35~mT) and for $T$ \textless30~K, weak four-fold symmetry patterns are found, also with intensities about 100 times weaker than for the ZFC case. They become stronger as the temperature decreases, resulting in a small pocket of a relatively low intensity in the contour plot of the scattered intensity of Fig. \ref{Fig:FC}(e). 


\begin{figure}
\includegraphics[width=0.8\linewidth]{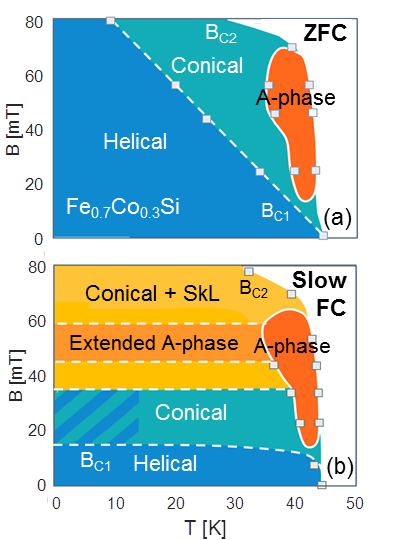}
\caption{\label{Fig:Phase_diagram} Proposed phase diagrams of Fe$_{0.7}$Co$_{0.3}$Si deduced from the SANS patterns and intensities for (a) Zero Field Cooling (ZFC) and (b) Slow Field Cooling (Slow FC) showing the helical, conical and $A$-phases. The slow FC diagram of panel (b) also shows the extended $A$-phase as well as the $B-T$ areas where helical or skyrmion lattice correlations coexist with the conical phase.}
\end{figure}

\section{\label{sec:disc} Discussion}
The results presented above lead to the phase diagrams of Fig. \ref{Fig:Phase_diagram}, which highlight the differences between ZFC and slow FC. The ZFC phase diagram reveals that the conical phase extends, as in the doped compound Mn$_{1-x}$Fe$_x$Si and Mn$_{1-x}$Co$_x$Si\cite{bauer2010} to much larger fields than in undoped compounds as MnSi and Cu$_2$OSeO$_3$ as well as that they reveal a temperature dependence of $B_{C1}$ for ZFC. The conical phase is greatly suppressed for the FC case where there are regions where metastable SkL or helical correlations coexist with the conical phase under field cooling conditions.

These SkL correlations exist over a very extended region of the phase diagram until the lowest temperature and are not only found at temperatures below the thermodynamically stable $A$-phase. The very low intensity, almost two orders of magnitude lower than at the $A$-phase, could possible indicate surface or edge pinning \cite{Du:2015dt, Rybakov:2015dh} that may stabilize these chiral correlations in directions perpendicular to the applied magnetic field. 

On the other hand, the existence of (isolated) biskyrmion and multiskyrmion states within the conical phase arising from an attractive interskyrmion potential has been established theoretically \cite{2016arXiv160202353L}. However, the results presented here rather support \textit{lattices} of skyrmions rather than \textit{isolated} skyrmions and are as such more in agreement with the theoretical computations of \cite{Wilson:2014gd}. They predict the stabilization of metastable skyrmion \textit{lattices} over a large fraction of the phase diagram below $T_C$. These metastable skyrmion lattice are formed by cooling through the precursor region above $T_C$ where they are nucleated. A subsequent drop in temperature below $T_C$ turns the skyrmion lattices in metastable states of which the stability increases with decreasing $T$ \cite{Wilson:2014gd}. 

These theoretical predictions are also inline with the differences observed between Fast and Field Cooling. During fast field cooling, the exposure of the skyrmion lattice to the region just below $T_C$ where the energy barrier heights are relatively small, is limited as compared to slow field cooling. This results in a smaller deterioration of the SkL correlations for the fast FC case and a stronger intensity at lower temperatures. 

A recent experimental study showed that metastable SkL correlations can also be quenched by applying extremely high cooling rates of $\sim$700 Ks$^{-1}$ in MnSi \cite{oike2016}. These high cooling rates are required to circumvent the unwinding of the SkL as observed in bulk Fe$_{0.5}$Co$_{0.5}$Si by \cite{milde2013} with magnetic force microscopy. However, the cooling rates applied in this study are more than three orders of magnitude higher, suggesting that the unwinding of the Skl in Fe$_{0.7}$Co$_{0.3}$Si occurs at a totally different timescale than in MnSi. The observation of these metastable skyrmion lattice phase over macroscopic time scales in Fe$_{0.7}$Co$_{0.3}$Si may be attributed to the combination of quenched chemical disorder that is due to the solid solution of Fe and Co.

\section{\label{sec:conclusion} Conclusion}
To conclude, we observe a pronounced history and cooling-rate dependence of the magnetic phase diagram below $T_C$ in Fe$_{0.7}$Co$_{0.3}$Si. By cooling under field, metastable skyrmion lattice correlations are observed outside the thermodynamically stable $A$-phase until the lowest temperature. These thermodynamic metastable skyrmion lattice correlations coexist with the conical phase and do not only appear at temperatures below the $A$-phase but also at fields smaller or higher than the fields required to stabilize the $A$-phase. The intensity of these skyrmion lattice correlations can be enhanced by increasing the cooling rate as the increased cooling limit possibly reduces the unwinding of the SkL in a region just below $T_C$.  The observation of these phenomena with the macroscopic cooling rates used in a neutron scattering experiment may be related to the quenched chemical disorder from the solid solution of Fe and Co.

\begin{acknowledgments}
LB, FQ and CP thank Maxim Mostovoy for valuable discussions and comments. The work of LB is financially supported by The Netherlands Organisation for Scientific Research through the project 721.012.102. FQ acknowledges financial support from China Scholarship Council (CSC). FQ and CP acknowledge funding from the European Union Seventh Framework Program [FP7/2007-2013] under grant agreement Nr. 283883. 
\end{acknowledgments}

\bibliography{FeCoSiSANS}


\end{document}